\pdfoutput=1

\documentclass[11pt]{article}

\usepackage[]{coling}

\usepackage{times}
\usepackage{latexsym}

\usepackage[T1]{fontenc}

\usepackage[utf8]{inputenc}

\usepackage{microtype}

\usepackage{inconsolata}
\usepackage{multirow}
\usepackage{graphicx}
\usepackage{amsmath}
\usepackage{pdfpages}
\usepackage{booktabs}

\usepackage{xcolor} 
\usepackage{colortbl} 
\PassOptionsToPackage{table}{xcolor}
\PassOptionsToPackage{dvipsnames}{xcolor}

\title{Multi-Facet Blending for Faceted Query-by-Example Retrieval}

 \author{Heejin Do$^{1}$, Sangwon Ryu$^{1}$, Jonghwi Kim$^{1}$, Gary Geunbae Lee$^{1,2}$ \\
  \centering
  \begin{tabular}[t]{c}
    $^{1}$Graduate School of Artificial Intelligence, POSTECH, South Korea \\
    $^{2}$Department of Computer Science and Engineering, POSTECH, South Korea \\
    \texttt{\{heejindo, ryusangwon, jonghwi.kim, gblee\}@postech.ac.kr} \\
  \end{tabular}
}

\begin{document}
\maketitle

\begin{abstract}
With the growing demand to fit fine-grained user intents, faceted query-by-example (QBE), which retrieves similar documents conditioned on specific facets, has gained recent attention. However, prior approaches mainly depend on document-level comparisons using basic indicators like citations due to the lack of facet-level relevance datasets; yet, this limits their use to citation-based domains and fails to capture the intricacies of facet constraints. In this paper, we propose a multi-facet blending (FaBle) augmentation method, which exploits modularity by \textit{decomposing} and \textit{recomposing} to explicitly synthesize facet-specific training sets. We automatically decompose documents into facet units and generate (ir)relevant pairs by leveraging LLMs' intrinsic distinguishing capabilities; then, dynamically recomposing the units leads to facet-wise relevance-informed document pairs. Our modularization eliminates the need for pre-defined facet knowledge or labels. Further, to prove the FaBle's efficacy in a new domain beyond citation-based scientific paper retrieval, we release a benchmark dataset for educational exam item QBE. FaBle augmentation on 1K documents remarkably assists training in obtaining facet conditional embeddings.

\end{abstract}

\section{Introduction}
\begin{figure}[t]
\centering
\includegraphics[width=\linewidth]{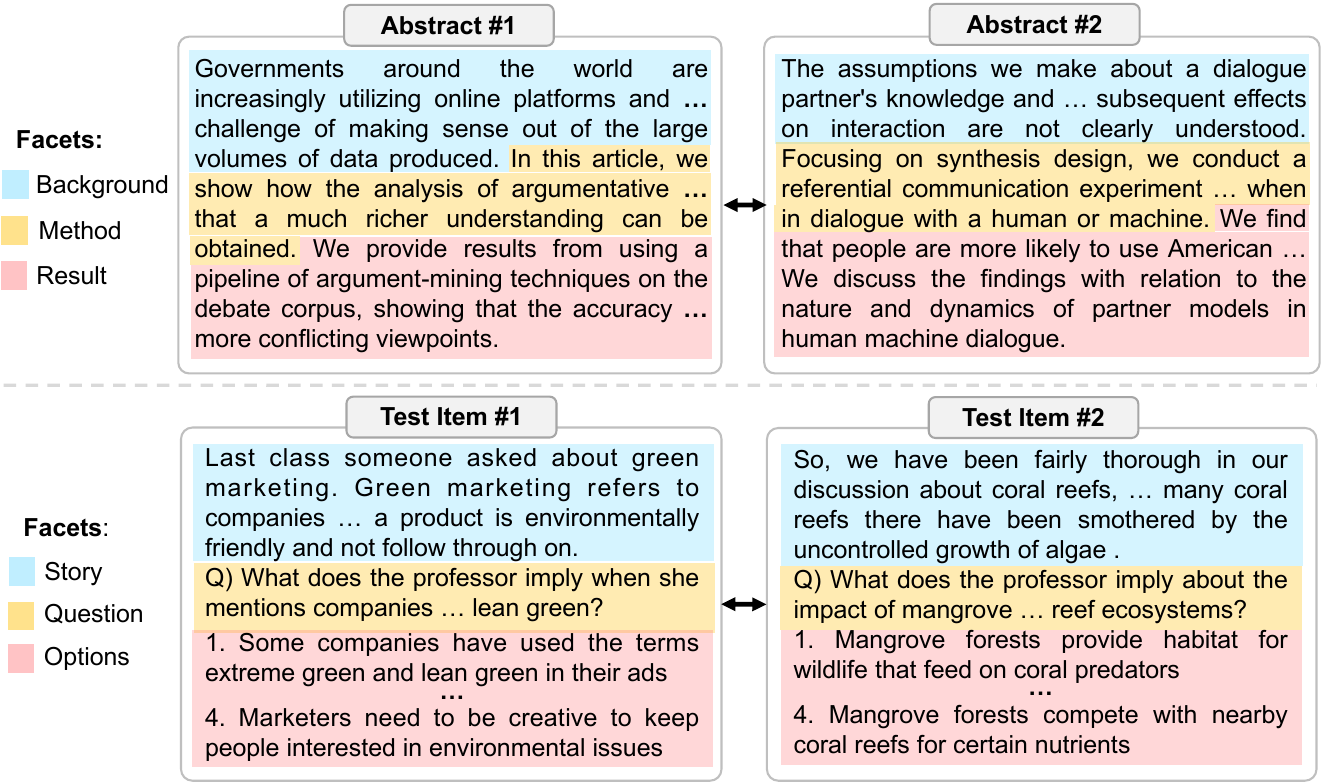}
\caption{Examples of documents with multiple facets.}
\label{fig: figure 0}
\end{figure}

Query-by-example (QBE), which involves retrieving relevant documents given a query document, is a fundamental technique in both exploratory search \cite{lissandrini2019example} and recommendation systems \cite{ostendorff-etal-2020-aspect, 10.1145/3383583.3398525, lee2013personalized}. However, documents typically include multiple facets distinguished by specific rhetorical units (e.g., \textit{background}, \textit{method}, and \textit{result} of academic paper abstract); thus, querying with the entire document, not identifying the specific facet of interest, can lead to unintentional or irrelevant retrievals (Figure~\ref{fig: figure 0}). For instance, to recommend exam items similar in question type to a student's incorrect answer, prioritize the \textit{question} facet for retrieval, regardless of \textit{story} or \textit{options}, is required.

Accordingly, faceted QBE, which conditions the query document on a specific facet, has garnered recent attention for intent-tailored fine-grained document search \cite{dunne2012rapid,hope-etal-2020-scisight, neves2019evaluation}. This task has been predominantly explored in scientific paper retrieval, relying on the vast amount of public corpora where citation labels provide superficial cues \cite{cohan-etal-2020-specter, ostendorff-etal-2022-neighborhood, mysore2021csfcube, mysore-etal-2022-multi}. However, those methods are not feasible for other domains (e.g., education or legal), where such citation labels are absent, and large-scale open-source corpora are lacking \cite{li2023thuir}.
Further, the reliance on document-level comparisons often leads to the failure to capture facet constraints, especially for intricate cases \cite{mysore2021csfcube}.


In this paper, we propose a multi-facet blending (FaBle) augmentation method, which dynamically exploits modularity with \textit{decomposing} and \textit{recomposing}. In particular, we first decompose each facet within the document by summary-driven identification, leveraging zero-shot prompting with sLLM. Then, we generate facet-wise \textit{similar} and \textit{dissimilar} facet fragments by self-feeding the decomposed facet summary in recursive prompting. Referring to the identified facet guides the synthesis of facet-aware compositions distinguished from other facets. Finally, \textit{recomposition} strategy integrates the synthesized facets to reconstruct facet-conditioned pseudo documents, creating positive--negative pairs for an anchor document. Fable explicitly create facet-specific training sets to assist model training for faceted QBE, eliminating the need for pre-defined facet knowledge or labels. 



We target scientific paper abstract retrieval for validation, as it is the sole field providing the benchmark test set for faceted QBE. Aiming to assist in a data-scarce scenario, we employ only 1K documents for augmentation without any citation labels.
Experimental results of fine-tuning the SPECTER \cite{cohan-etal-2020-specter} model with FaBle-augmented pairs are comparable or better to previous models, where more than 1.3M training sets were used for fine-tuning. Notably, FaBle significantly improves the challenging \textit{method} facet, even outperforming the strong prior models. This result highlights that our fine-grained augmentation overcomes the limitations of coarse-grained approaches that ill-captured intricate facets \cite{mysore2021csfcube}.

To further evaluate FaBle's domain scalability and practical efficacy, we present a novel test set for faceted educational exam item retrieval, FEIR, derived from the TOEFL-QA data. Applying FaBle to educational items remarkably improves performance across all facets, demonstrating domain-agnostic effects.
We expect the FEIR to stimulate future works of faceted QBE in this emerging education domain. Codes and datasets are on GitHub\footnote{https://github.com/doheejin/FaBle}.

\begin{figure*}[t]
\centering
\includegraphics[width=0.93\linewidth]{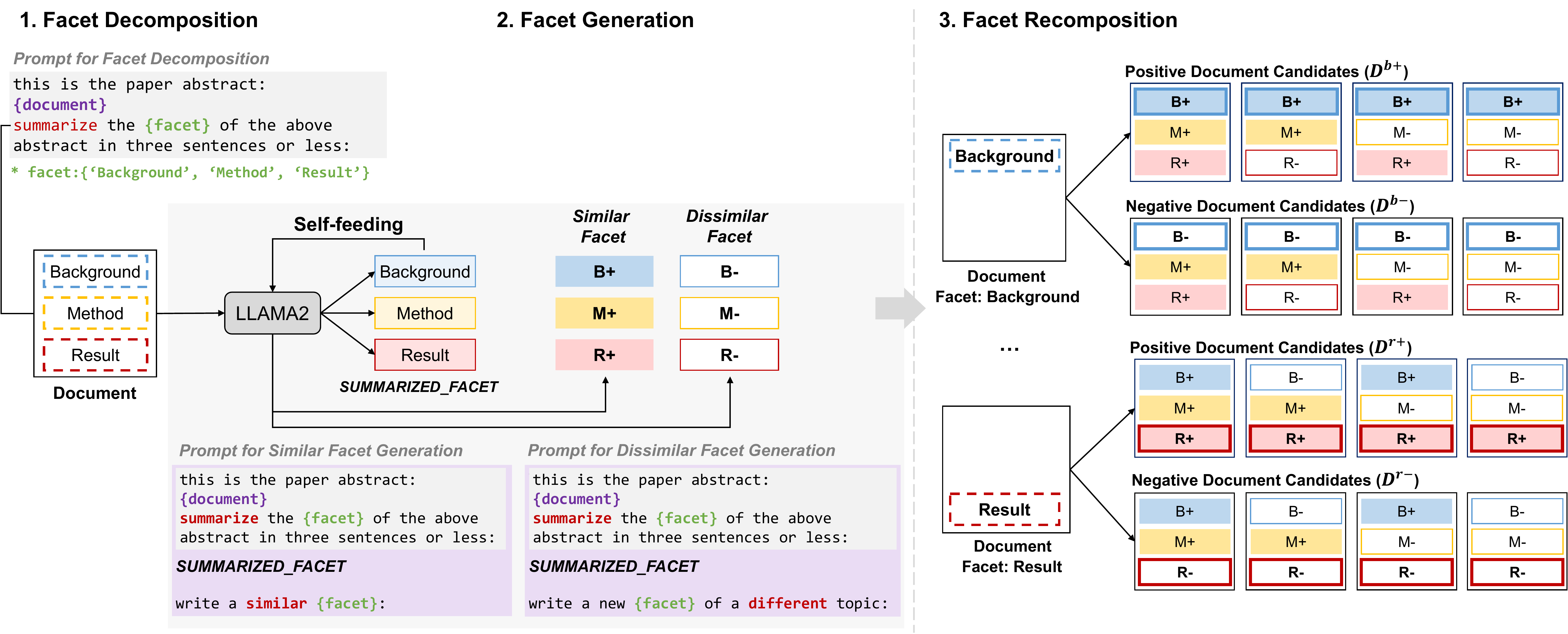}
\caption{The overview of the FaBle method and examples of detailed prompts used for scientific paper retrieval. 
}
\label{fig: figure 1}
\end{figure*}


\section{Related Work}

\paragraph{\textbf{QBE}} 
QBE is a fundamental task across diverse fields, such as legal or academic, where document-level findings for recommendation or exploratory search are important \cite{lissandrini2019example, ostendorff-etal-2020-aspect, 10.1145/3383583.3398525, lee2013personalized}. Most prior studies focused on retrieving scientific papers, using large-scale datasets and estimating similarities based on citations \cite{cohan-etal-2020-specter, mysore2021csfcube, mysore-etal-2022-multi, ostendorff-etal-2022-neighborhood}. \citet{cohan-etal-2020-specter} introduced the SPECTER to obtain document-level embeddings by measuring similarity via citation graphs, and \citet{ostendorff-etal-2022-neighborhood} used a citation embedding graph combined with neighbor contrastive learning.


\paragraph{\textbf{Faceted QBE}}
Documents typically encompass multiple facets; thus, considering overall document-level relevance may not align with user intent \cite{do2024aspect}. Faceted QBE has emerged to address this, enabling facet-level document comparisons \cite{neves2019evaluation, doris-mae}. While most studies focus on scientific paper retrieval \cite{mysore2021csfcube, mysore-etal-2022-multi, doris-mae}, they do not directly train on facet-wise relevance annotated data, as such data is difficult to obtain. 
Instead, \citet{mysore2021csfcube} utilized an additional 66K citation-based pair for training, and \citet{mysore-etal-2022-multi} used 2.6M co-citation sentences with an auxiliary optimal transport technique. However, the reliance on abundant domain-specific data and citations restricts their applicability to other low-resource domains.

\paragraph{\textbf{LLM Augmented Retrieval}} 
LLM-based augmentation techniques have evolved from using GPT-2 \cite{radford2019language} to GPT-3 \cite{brown2020language} models to address the lack of relevance annotations. \citet{luu-etal-2021-explaining} fine-tune GPT-2 to generate relationships between two scientific papers, assuming in-text citation sentences elucidate their connections. \citet{gao-etal-2023-precise} use GPT-3 to generate hypothetical documents corresponding to desired instructions in a zero-shot manner. 

Recently, for faceted QBE, \citet{doris-mae} utilize ChatGPT to annotate the relevance scores of aspect-paper pairs, reducing the burden of human labor. Despite aiming at sub-aspect level similarity evaluation, utilizing ChatGPT for massive datasets still incurs significant costs; thus, they mainly target \textit{testing} faceted QBE, not \textit{training}. Also, as they only contain computer science-related documents, datasets are not generalizable to other fields. Contrarily, by leveraging the capacity of open-source smaller LLM, we eliminate the cost burden and introduce the domain-extendable method.



\section{FaBle: Multi-facet Blending}
For general QBE, obtaining informative representations for query and candidate documents is crucial to effectively retrieve similar documents. To achieve this, model training requires a triplet pair ($D^Q$, $D^+$, $D^-$) comprising a query document, a positive document, and a negative document. In faceted QBE, queries include additional facet conditions; thus, facet-constrained triplet pairs can lead to more precise and focused model training. Unlike prior methods that implicitly construct $D^+$ and $D^-$ based on citations on $D^Q$, we explicitly construct facet-conditional triplet pairs ($D^{f; Q}$, $D^{f+}$, $D^{f-}$). 

FaBle mainly comprises three stages (Fig.~\ref{fig: figure 1}): decomposition (§3.1), generation (§3.2), and recomposition (§3.3). 
In this section, we explain examples of scientific paper retrieval, but FaBle is broadly applicable to domains with distinct facets.

\subsection{Facet Decomposition}
\begin{figure*}[]
\centering
\includegraphics[width=0.87\textwidth]{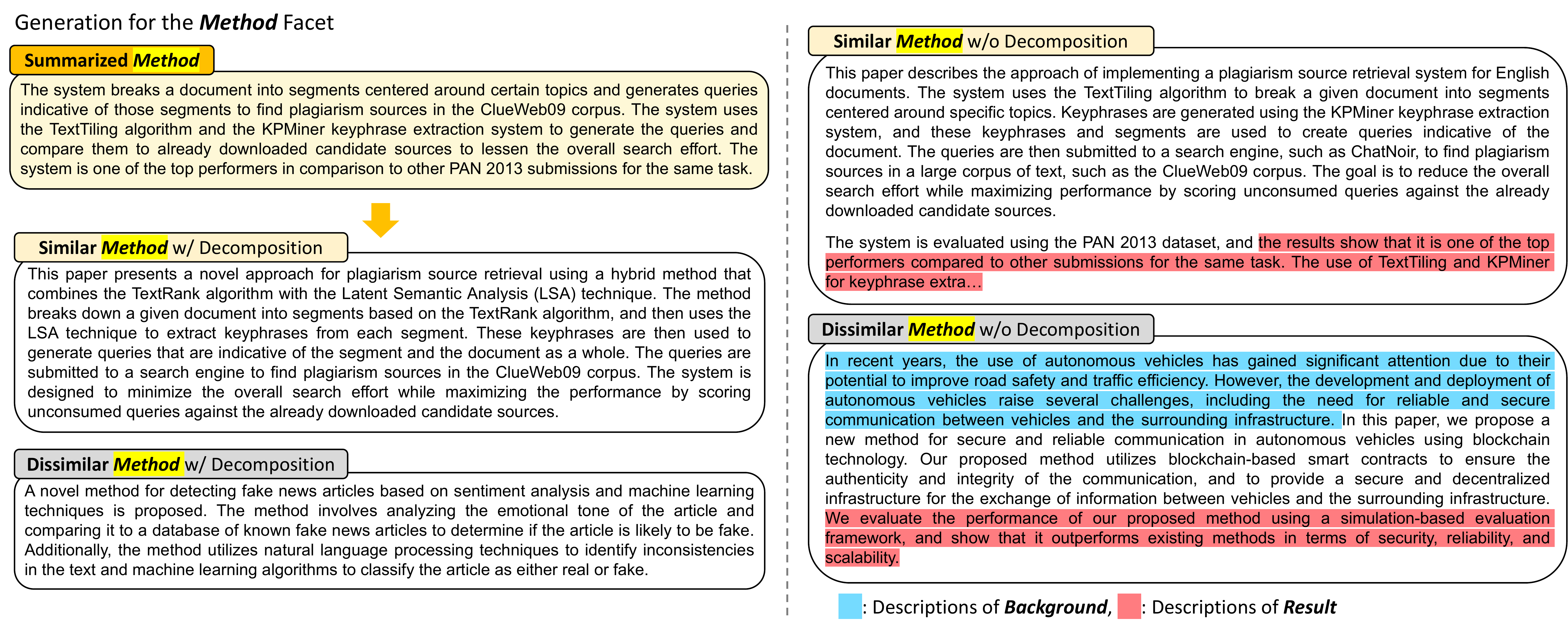}
\caption{Examples of the generated \textit{similar} and \textit{dissimilar} \textit{method} facets with our self-fed decomposition (left) and without the (\textit{w/o}) decomposition (right). Directly generating \textit{similar} and \textit{dissimilar} facets without decomposition can lead to the results containing facets other than the intended one, as highlighted.}
\label{fig: figure 2}
\end{figure*}

To identify each facet, we first decompose the document into multiple facet units. For this, we prompt LLM to summarize a specific facet in a zero-shot manner. We use the publicly available sLLM, LLaMA2-13B \cite{touvron2023llama}, taking advantage of open and easy access. By prompting the model to summarize a desired facet within the document, the intended facet-distinct information is extracted. Given a document $D$, summarization prompt $p_{sum}$, and a facet name $f$, where $f \in \{\textit{background}, \textit{method}, \textit{result}\}$, as input, the model generates facet summary $S^f$, which modularize the $f$ facet: $S^f = \mathtt{Model}(D, p_{sum}, f)$. Figure~\ref{fig: figure 1} describes the detailed prompt, and Figure~\ref{fig: figure 2} shows an output summary example. The generated summary highly represents the facet, but it does not mean a \textit{real facet}; instead, it serves as an indicator to guide the subsequent generation stage.

\subsection{Facet Generation}
To generate each facet-specific similar and dissimilar fragment, the same model self-fed the prior prompt used to decompose and its extracted output as shown in Figure~\ref{fig: figure 1}. 
Although LLaMA2 has proved proficiency in various generation tasks, its zero-shot performance often lags behind task-specific instruction tuning or GPT-4 \cite{zhu2023endtoend, openai2023gpt4}. Our self-feeding approach aids in target-oriented generation by referring to the facet-identified summary while eliminating the burden of fine-tuning. In particular, to generate $f$-facet similar component $C^f_{sim}$, the model takes pre-generated summary $S^f$ and the similar-generation prompt $p_{sim}$ as the input. For \textit{dissimilar} component $C^f_{dis}$, the model takes summary $S^f$ and dissimilar-generation prompt $p_{dis}$ as input:
\begin{eqnarray}
C^f_{sim} &=& \mathtt{Model}(D, p_{sum}, f, S^f, p_{sim}) \\
C^f_{dis} &=& \mathtt{Model}(D, p_{sum}, f, S^f, p_{dis}) 
\end{eqnarray}

Figure~\ref{fig: figure 2} reveals that our two-stage approach results in more target-facet-focused texts (left), while the simple prompting without the facet-identified summary outputs non-target facets mixed in (right). 

\subsection{Facet Recomposition}
To obtain the negative and positive document pairs for a query document conditioning a specific facet, we combine the generated \textit{similar} and \textit{dissimilar} facet components with a suitable recomposition recipe. The $f$-facet conditional positive $D^{f+}$ and negative $D^{f-}$ documents with total $n$ facets are:
\begin{eqnarray}
D^{f+} &=& [C^f_{sim}; C^{f_i\in F-f}_{sim|dis};\dots; C^{f_n\in F-f}_{sim|dis}] \\
D^{f-} &=& [C^f_{dis}; C^{f_i\in F-f}_{sim|dis};\dots; C^{f_n\in F-f}_{sim|dis}]
\end{eqnarray}
where [$;$] denotes concatenation, $F$ is a set of facets, and $f_i\in F-f$ is a facet different from the target facet $f$. 
Consequently, the triplet pair $(D^{f;Q}, D^{f+}, D^{f-})$ is constructed for the query document $D^{f;Q}$, conditioned on a target facet $f$.


On a single original document with three facets, four $D^{f+}$ and four $D^{f-}$ are generated via facet recomposition. Then, five documents, including the original one, lead to ten $(D^{f;Q}, D^{f+})$ pairs (i.e., five choose two, $\binom{5}{2}$). For each of them, one $D^{f-}$ is selected among four candidates, resulting in a total of forty $(D^{f;Q}, D^{f+}, D^{f-})$ pairs per sample. Note that FaBle operates without any labels, including weak labels like citations or pre-divided facet tags.

\subsection{Fine-tuning for Faceted QBE}
\label{sec: triplet}
We validate the efficacy of FaBle-augmented triplet pairs in model training via contrastive learning, the widely adopted mechanism for representation learning. Specifically, we employ a pre-trained SciBERT \cite{beltagy-etal-2019-scibert}-based SPECTER \cite{cohan-etal-2020-specter} to embed the documents. We fine-tune the model with triplet loss to verify whether the synthesized dataset benefits model training. Our loss function $L(D^{f;Q}, D^{f+}, D^{f-})$ is defined as:
$max \left\{ (\mathrm{d}(D^{f;Q},D^{f+})-\mathrm{d}(D^{f;Q},D^{f-})+m),0\right\}$
where $\mathrm{d}$ is a distance function, and $m$ is the loss margin hyperparameter. Note that no additional modeling techniques are used to examine the unique effects of the augmentation.

\begin{figure}[]
\centering
\includegraphics[width=0.9\linewidth]{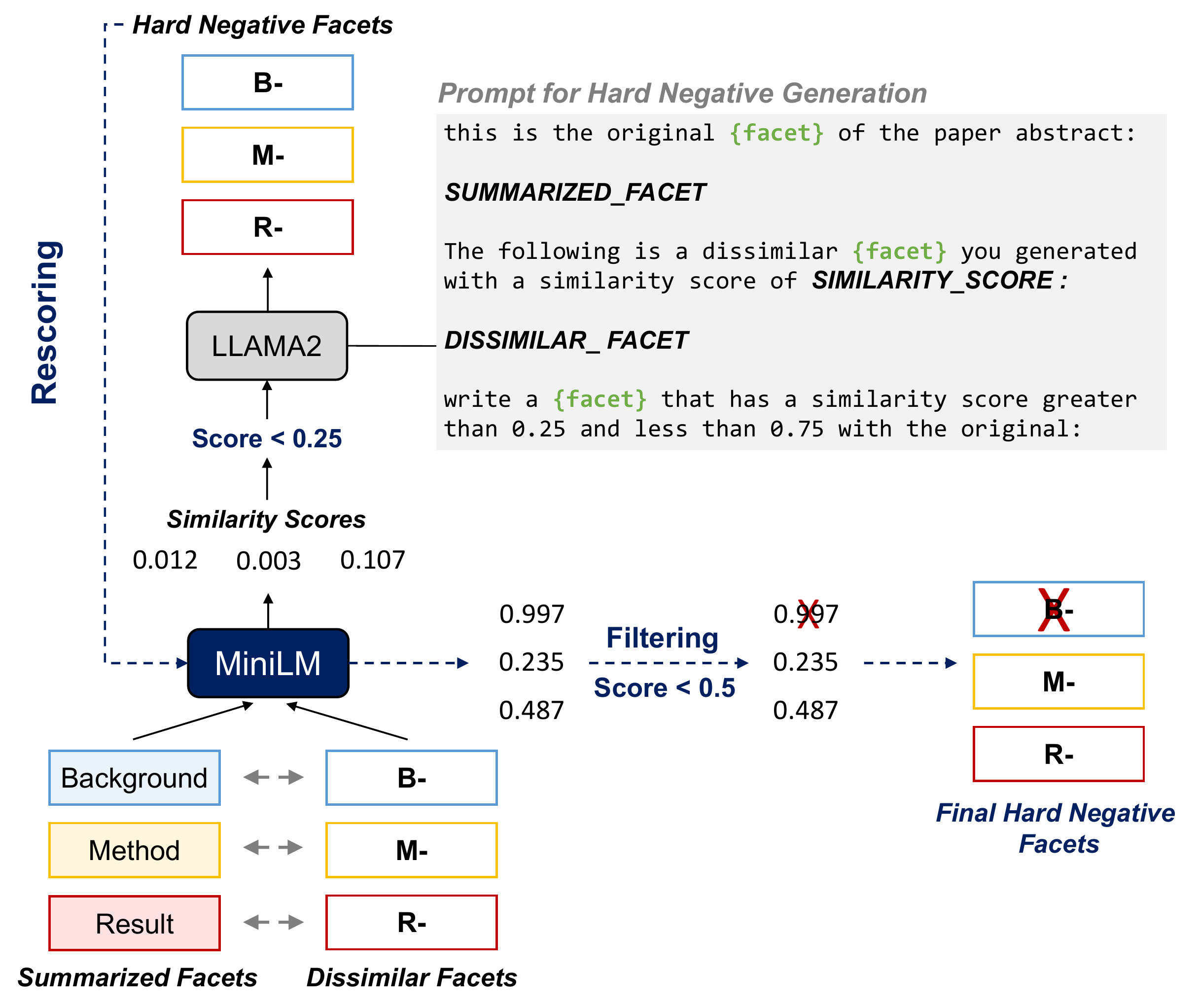}
\caption{Hard negative generation procedure (§~\ref{sec3.5}). 
}
\label{fig: figure 4}
\end{figure}

\subsection{Hard Negative Generation} \label{sec3.5}
The significance and efficacy of hard negative mining for retrieval tasks have been widely demonstrated \cite{xiong2020approximate, zhan2021optimizing, zhang2021adversarial, zhou-etal-2023-towards-robust}. These studies highlight that more challenging negative samples lead to better representation capturing. In this work, we explicitly prompt the LLM to create facets of different topics to generate negative (dissimilar) ones for a specific facet. This may compel the generation of easily distinguishable snippets, potentially leading to the absence of hard negative samples.

Thus, to enhance the FaBle-generated facets from the perspective of negative sampling, we employ MiniLM\footnote{https://huggingface.co/cross-encoder/ms-marco-MiniLM-L-6-v2}\cite{wang2020minilm}, a lightweight cross-encoder model trained on MS MARCO \cite{bajaj2016ms} using knowledge distillation, after Stage 2 (Figure~\ref{fig: figure 4}). With its proven high performance and efficient inference time \cite{thakur2021beir}, MiniLM is ideal for pseudo-relevance scoring. The similarity score, $\mathtt{MiniLM}(S^f, C^f_{dis})$, is measured with the summarized facet $S^f$ and the generated \textit{dissimilar} facet component $C^f_{dis}$ inputs.
The output score reflects how closely the generated facet fragments align with the original facets. Based on the score distribution, we regard the negative samples with a similarity score below 0.25 as \textit{easy negatives}. Here, we aim to regenerate those samples to have a specific score distribution of 0.25--0.5 for hard negative mining. To control the relevance level, we notify the LLM of the current similarity score by including it in the prompt, inspired by recent studies that incorporate exact numeric values in instructions \cite{ribeiro2023generating, zhang2024benchmarking}. We then measure the MiniLM scores for the regenerated facets and identify those below 0.5 as hard negatives. The recomposition process in Stage 3 is applied to the added facets, yielding the final supplemental hard negatives.

\begin{table}[]
\centering
\scalebox{0.63}{\begin{tabular}{l|ccc|cc}
\hline
&\multicolumn{3}{c|}{Orig} & \multicolumn{2}{c}{FEIR} \\\cline{2-6} 
    & Train & Valid & Test & Query & Cands\\ \hline
Full & 717     & 124 & 122  &  - & - \\
\hline
Story & 150     & 24 & 24 & 8 & 23 \\
Question & 717     & 124 & 122 & 8 & 80 \\
Options & 717     & 124 & 122 & 8 & 70 \\
\hline
\end{tabular}}\caption{Summary of the original TOEFL-QA dataset (\textit{Orig}) and the FEIR test set. \textit{Story} is a shared facet among multiple question-options sets. 
}
\label{tab1}
\end{table}

\section{FEIR}


The benchmark test set for faceted QBE is absent in domains other than scientific paper retrieval. This gap leads to a shortage of related studies in other fields, such as educational item retrieval, where each item comprises multiple facets. Even when items share similar \textit{Questions}, their \textit{Stories} and \textit{Options} may differ, requiring fine-grained search queries. To validate the scalability of FaBle and support future research, we introduce a Faceted Educational exam Item Retrieval (FEIR) test set for the underexplored language education domain.

\paragraph{Dataset Construction}
We employ exam items from the publicly available TOEFL-QA\footnote{https://github.com/iamyuanchung/TOEFL-QA/}\cite{chung2018supervised, tseng2016towards} dataset, a representative English as a Foreign Language (EFL) exam, to build the FEIR. The dataset contains 963 TOEFL listening QA items, and we utilize 122 test set items for constructing the FEIR test set (Table~\ref{tab1}). 
Inspired by CSFCube \cite{mysore2021csfcube}, which has 16 queries per facet, and given our limited original dataset, we form 8 query items for each facet (total 24 queries). To ensure diversity in relevance scores, we evaluated each sample's similarity with MiniLM scores and sequentially selected eight unique samples with the largest standard deviations in their score distributions. Each facet contains four conversation-type and four lecture-type queries. 
For candidate selection in the \textit{story} facet, where data is limited, we use all 23 remaining items except the query item. In the \textit{question} and \textit{options} facets, we choose 80 and 70 items, prioritizing those with the highest standard deviations after removing the query items.


\begin{figure}[]
\centering
\includegraphics[width=\linewidth]{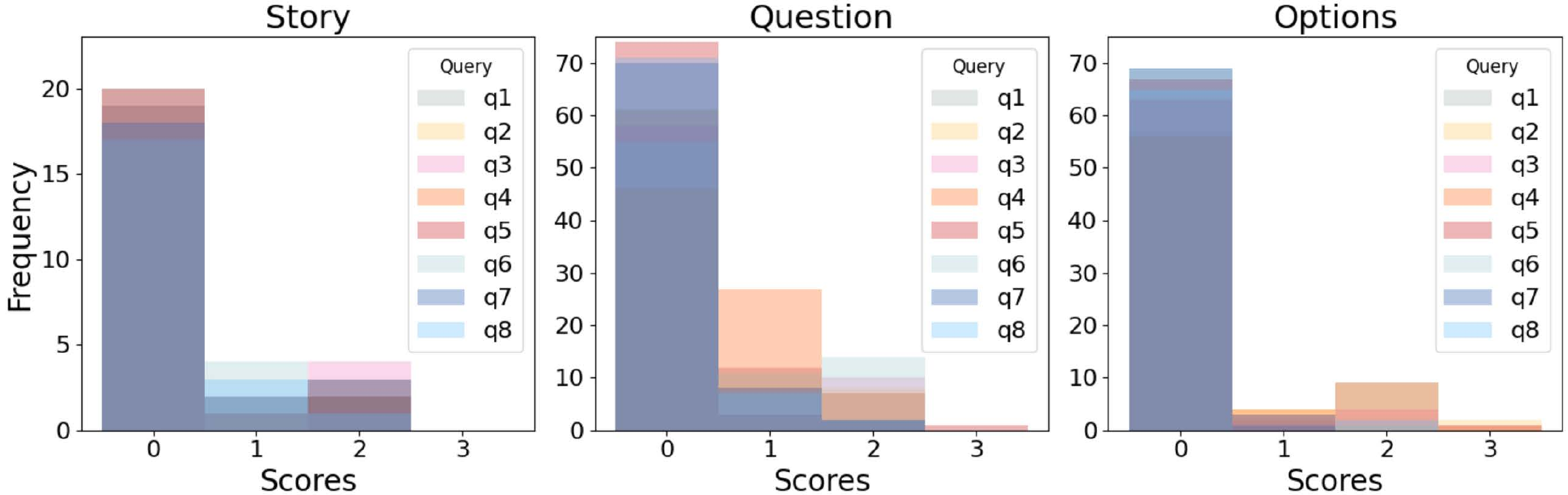}
\caption{Score label distributions per query by facet. } 
\label{fig: fig5}
\end{figure}

\begin{table*}[t]
\centering
\scalebox{0.67}{
\begin{tabular}{l|cc|cc|cc|cc}
\hline
CFSCUBE Facets  & \multicolumn{2}{c}{Background} & \multicolumn{2}{c}{Method} & \multicolumn{2}{c}{Result} & \multicolumn{2}{c}{Aggregated} \\ \cline{2-9} 
Model & NDCG$_{\%20}$ & MAP & NDCG$_{\%20}$ & MAP & NDCG$_{\%20}$ & MAP & NDCG$_{\%20}$ & MAP \\ \hline
\texttt{SentBERT-PP}  & 60.80 &  -  & 33.40 &  -  & 52.35 & -    & 48.57 &  - \\
\texttt{SentBERT-NLI} & 54.23 &  -  & 31.10 &  -  & 51.30 & -    & 45.39 &  - \\
\texttt{CoSentBert}  & 61.27 & 35.78  & 38.77 & 19.27  & 50.68 & 32.15   & 50.68 & 28.95 \\
SCINCL   & 70.02 & 49.64  & 46.61 & 27.14  & 61.70 & 41.83  & 59.24 & 39.37 \\ 
SPECTER-ID & 69.22 & -   & 42.76 & - & 60.40 & - & 57.22 & -  \\
TSASPIRE$_{\texttt{Spec}}$  & 70.22 & 49.58   & 48.20 & 28.86  &  64.39 & 42.92  & 60.71 & 40.26 \\
OTASPIRE$_{\texttt{Spec}}$  & \underline{71.04} & 50.56 &  46.46 & 27.64  & \underline{67.38} & \underline{44.75} & \underline{61.41} & \underline{40.79}  \\
TS+OTASPIRE$_{\texttt{Spec}}$ & 70.99 & \underline{51.79} & 47.60 & 26.68   & 64.82 & 43.06  & 60.86 & 40.26 \\ \hline
SPECTER    & 66.70 & \textbf{43.95}  & 37.41 & 22.44  & 56.67 & 36.79  & 53.28 & 34.23  \\
+FaBle (Ours)      & \textbf{67.38} & 42.66 & \textbf{44.97} & \textbf{25.98} & \textbf{58.10} & \textbf{38.60}  & \textbf{56.60} & \textbf{35.60} \vspace{-0.3em}\\ 
 & {\small±0.28} & {\small±0.32}	& {\small±0.16} & {\small±1.05} & {\small±1.78} & {\small±1.31} & {\small±0.57} & {\small±0.52} \\
\hline
SPECTER-COCITE$_{\texttt{Scib}}$ & 68.71 & 48.40  & 46.79 & 26.95  & 59.68 & \textbf{38.93}  & 58.16 & 37.90 \\
SPECTER-COCITE$_{\texttt{Spec}}$ & 70.03 & \textbf{49.99}   & 45.99 & 25.60  & 59.95 & 37.33  & 58.38 & 37.39 \\
+FaBle$_{\texttt{Spec}}$ (Ours) & \textbf{70.09} & 45.93 & 49.14 & 30.90 & {60.88} & {38.08} & \textbf{59.79} & 38.11 \vspace{-0.3em}\\ 
& {\small±0.09} & {\small±0.54}	& {\small±0.95} & {\small±0.89} & {\small±0.86} & {\small±0.20} & {\small±0.26} & {\small±0.31} \\
+FaBle$_{\texttt{Spec}}$+HN (Ours)  & 69.48 & 46.03 & \underline{\textbf{49.43}} & \underline{\textbf{32.57}} & \textbf{61.09} & {38.14} & {59.76}& \textbf{38.73} \vspace{-0.3em}\\
& {\small±0.83} & {\small±0.60}	& {\small±1.11} & {\small±1.32} & {\small±0.37} & {\small±0.64} & {\small±0.75} & {\small±0.66} \\
\hline
\end{tabular}}\caption{Evaluation results on CSFCube test set. \textit{SPECTER-COCITE$_{\texttt{Spec}}$} and \textit{SPECTER-COCITE$_{\texttt{Scib}}$} are the SPECTER- and SciBERT \cite{beltagy-etal-2019-scibert}-initialized model trained with co-citation dataset, respectively. \textit{+FaBle} and \textit{+FaBle$_{\texttt{Spec}}$} denote fine-tuning on the above \textit{SPECTER} and \textit{SPECTER-COCITE$_{\texttt{Spec}}$}, respectively. \textit{+FaBle$_{\texttt{Spec}}$+HN} is the addition of \textbf{H}ard \textbf{N}egative samples. \textbf{Bold}: the highest among baseline and proposed methods, \underline{underline}: the highest score in each column, ±: standard deviation of three runs.}\label{table 1}
\end{table*}

\paragraph{\textbf{Relevance Annotation}}
To annotate relevance between facet-specific query-candidate pairs, we hired three experts: a professor in language learning major and two English specialists from Upwork\footnote{
https://www.upwork.com/}. Each facet was assigned to two different experts. Following detailed guidelines (Appendix~\ref{appendix3}), they rated the relevance of each query and candidate item on a 0--3 scale, similar to \citet{mysore2021csfcube}. The rounded average of two ratings is the final score. Figure~\ref{fig: fig5} shows the score distribution of candidates per query, with a minority being labelled between 1 and 3. This trend mirrors the CSFCube test set, where an average of 36.9 candidates per query are rated 1, and 9.8 candidates receive scores of 2 or 3. We examine the inter-annotator agreement by measuring the correlations between two annotators' labels: Kendall's $\tau$, Spearman's $\rho$, and Pearson's $r$. The facet-average values are 0.474, 0.492, and 0.557, respectively (p<0.05), indicating positive agreements \cite{chiang2023can}.




\section{Experiments}
\paragraph{Data and Settings}
We use only 1017 random paper abstracts from the 81.1M papers in the open-source S2ORC\footnote{https://allenai.org/data/s2orc} corpus \cite{lo-etal-2020-s2orc}, having metadata, abstracts, and full text of academic papers. However, we do not use any annotated information in this work. By deliberately limiting the initial data to a small amount (approximately 0.00125\%), we aim to validate that our method is effective in practical data-scarce settings. As the CSFCube comprises scientific papers in the computer science domain, we also select abstracts from the same field. Applying the FaBle with 1K documents, 40 triplet document pairs are generated per facet for a single document, resulting in 40.68K triplet pairs. To apply FaBle for education exam items, we use 717 items from the TOEFL QA training set, creating total 28.68K pairs 
As the dataset already has facet labels, we directly employ Stages 2 and 3. Detailed settings are in Appendix~\ref{appendix1}.

\begin{table}[t]
\centering
\scalebox{0.7}{%
\begin{tabular}{l|ccc}
\hline
    & Background & Method & Result \\ \hline
Summarized Facet ($S^f$)& \textbf{0.756}     & 0.668 & 0.685 \\
Similar Component ($C^f_{sim}$) & \textbf{0.736}     & 0.634 & 0.649 \\ \hline
\end{tabular}%
} \caption{Averaged similarity scores between the entire document and each facet (denoted as $S^f$ and $C^f_{sim}$).}\label{table 2}
\end{table}






\begin{table*}[t]
\centering
\scalebox{0.65}{
\begin{tabular}{l|cc|cc|cc|cc}
\hline
CFSCUBE Facets  & \multicolumn{2}{c}{Background} & \multicolumn{2}{c}{Method} & \multicolumn{2}{c}{Result} & \multicolumn{2}{c}{Aggregated} \\ \cline{2-9} 
Model  & NDCG$_{\%20}$ & MAP & NDCG$_{\%20}$ & MAP & NDCG$_{\%20}$ & MAP & NDCG$_{\%20}$ & MAP \\ \hline
SPECTER-COCITE$_{\texttt{Spec}}$ & 70.03 & \textbf{49.99}   & 45.99 & 25.60  & 59.95 & 37.33  & 58.38 & 37.39 \\
+FaBle$_{\texttt{Spec}}$      & \textbf{70.09} & 45.93 & \textbf{49.14} & \textbf{30.90} & \textbf{60.88} & \textbf{38.08} & \textbf{59.79} & \textbf{38.11}  \\ 
+FaBle-RN$_{\texttt{Spec}}$    & 69.61 & 46.58 & 46.82 & 28.62 & 59.83 & 37.47 & 58.48 & 37.32 \\
\hline
\end{tabular}}\caption{Ablation study results. While \textit{FaBle} includes the generation of \textit{dissimilar} facets in Stage 2, \textit{FaBle-RN} selects Random facets as Negatives. \textit{+FaBle$_{\texttt{Spec}}$} and \textit{+FaBle-RN$_{\texttt{Spec}}$} denote fine-tuning on the above model.}\label{table 3}
\end{table*}



\paragraph{Baselines}
Most studies on faceted QBE have used or fine-tuned the SPECTER \cite{cohan-etal-2020-specter} model; hence, we adopt it as our baseline. Our primary aim is to evaluate the efficacy of facet-specific augmentation in data-scarce settings rather than resorting to supplementary methods for fine-grained QBE. Thus, our comparisons focus on the baseline models and those fine-tuned with FaBle-augmented data. 
We train two versions: the original SPECTER and SPECTER-COCITE$_\texttt{SPEC}$. The latter is similar to SPECTER but was additionally trained on 1.3M co-citation datasets from \citet{mysore-etal-2022-multi} with 2--3 point aggregation across queries. We also assess whether the FaBle-assisted model is comparable to other strong models for faceted QBE, with further details in Appendix~\ref{appendix2}.

\paragraph{\textbf{Evaluation}} For evaluation, we use CSFCube\footnote{https://github.com/iesl/CSFCube} \cite{mysore2021csfcube} test set, which provides annotations for faceted QBE on computer science papers. 50 query abstract--facet pairs are assigned relevance scores (0--3). 
We use the FEIR set to evaluate the educational exam item. For metrics, we employ normalized discounted cumulative gain at rank K (NDCG@K) and mean average precision (MAP). In particular, we report NDCG$_{\%20}$, computing at 20\% of the query pool size, following prior works \cite{wang2013theoretical, mysore2021csfcube, mysore-etal-2022-multi}. For the FEIR with fewer queries and candidates, we also report the NDCG$_{\%10}$.


\section{Results}

Table~\ref{table 1} shows the main results of FaBle across three facets. Incorporating FaBle with SPECTER enhances performance in all facets, yielding notable average gains of 3.4\% in NDCG$_{\%20}$ and 1.4\% in MAP. For SPECTER-COCITE, fine-tuning the model with FaBle also improves the performance, highlighting our assistance in model training.


\paragraph{\textbf{Facet-Specific Results}}


\begin{figure}[t]
\centering
\includegraphics[width=0.97\linewidth]{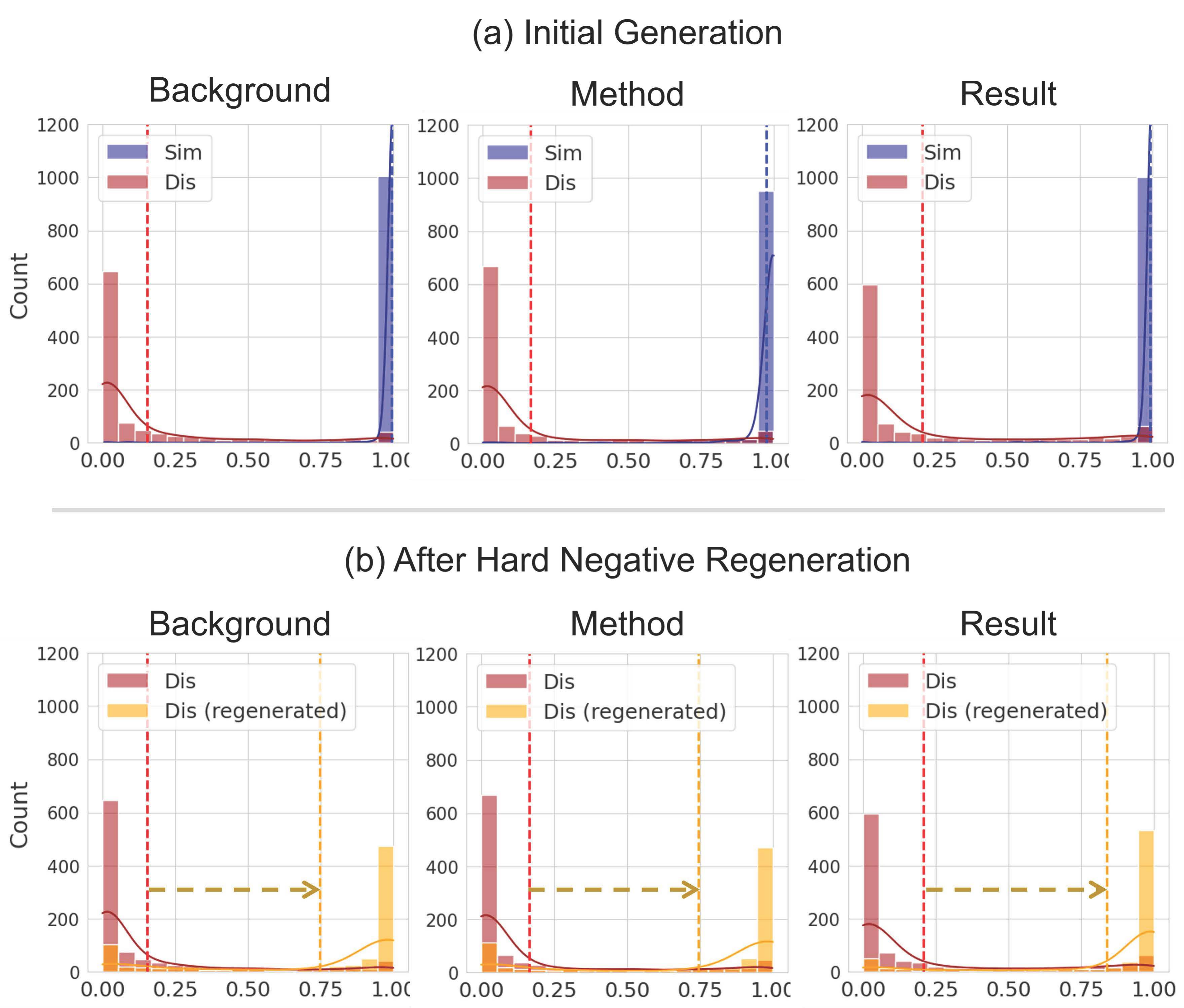}
\caption{MiniLM score distributions of generated \textit{Sim} and \textit{Dis} facets. (a) Initial distributions; (b) shifts in negative samples after regeneration; dashed line: mean.}
\label{fig: figure 5}
\end{figure}

The \textit{method} facet, widely recognized as the most challenging primarily due to its focus on procedural descriptions of technical concepts, encountered difficulties in assessing similarity with prior models \cite{mysore2021csfcube, mysore-etal-2022-multi}. In this context, the remarkable enhancements in the \textit{method} facet are noteworthy: an increase of 7.6\% in NDCG$_{\%20}$ and 3.5\% in MAP scores over SPECTER. Moreover, FaBle with the SPECTER-COCITE achieved a 3.4\% rise in NDCG$_{\%20}$ and a 7.0\% increase in MAP scores, even outperforming the robust ASPIRE models, trained on $\approx$32 times greater dataset than FaBle and employ co-citations labels with additional optimal transport techniques. Unlike them, FaBle leverages the knowledge embedded within LLMs trained on massive corpora to make intrinsic judgments about \textit{similarity} by individual facets. This enables the generation of sentences that deliberately mirror or distort procedural domain knowledge, resulting in sophisticated candidate construction, even for complex facets. Thus, the synthesized data can contribute to more discriminative representations for retrieval.



However, the \textit{background}, already achieved high scores (52.27\% higher NDCG$_{\%20}$ than the \textit{method} on SPECTER-COCITE$_{\texttt{Spec}}$), shows modest improvements, with a slight decline in MAP. This outcome is attributable to the comprehensive nature of the paper's background \cite{andrade2011write}, which existing coarse-grained systems can sufficiently capture; small gaps across all models also support this. Further, \citet{mysore-etal-2022-multi} noted that the stronger correlation between background contents and the paper's overall topic leads to the success of general models that incorporate whole abstract-level representations. 
We find this dependency by examining similarity scores between each decomposed and generated facet and full document. Table~\ref{table 2} reveals that the \textit{background} facet has a higher similarity to the entire document than other facets, explaining the less pronounced impact of our facet-specific approach on it.


\begin{figure}[t]
\centering
\includegraphics[width=0.96\linewidth]{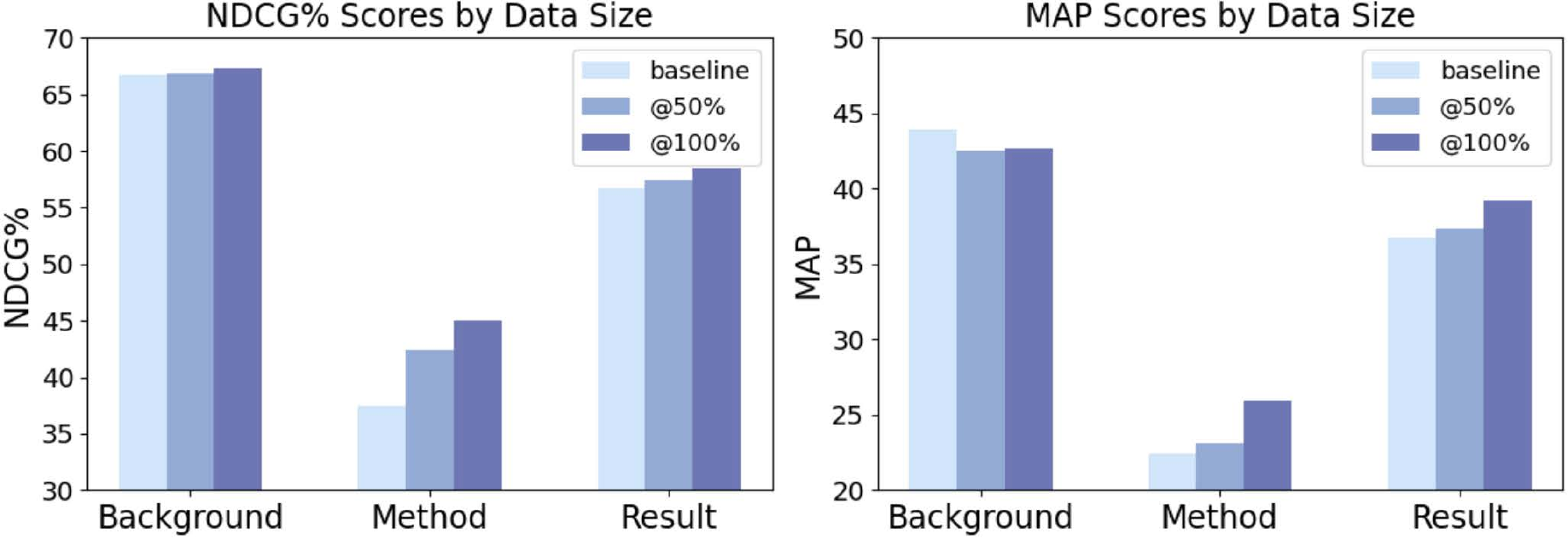}
\caption{Comparison of NDCG$_{\%20}$ (left) and MAP (right) performances by the dataset size per facet.}
\label{fig: figure 7}
\end{figure}


\begin{table*}[t]
\centering
\scalebox{0.64}{
\begin{tabular}{l|ccc|ccc|ccc|ccc}
\hline
 & \multicolumn{3}{c}{Story} & \multicolumn{3}{c}{Question} & \multicolumn{3}{c}{Options} & \multicolumn{3}{c}{Aggregated} \\ \cline{2-13} 
Model  & NDCG$_{\%20}$ & NDCG$_{\%10}$ & MAP & NDCG$_{\%20}$ & NDCG$_{\%10}$ & MAP & NDCG$_{\%20}$ & NDCG$_{\%10}$ & MAP & NDCG$_{\%20}$ & NDCG$_{\%10}$ & MAP \\ \hline
SCINCL & \textbf{69.15} & 71.88 & 60.79 & 29.64 & 23.05 & 19.00 & \textbf{80.26} & \textbf{78.81} & 58.67 & 59.68 & 57.91 & 46.15 \\ 
{+FaBle} & 69.11 & \textbf{76.04} & \textbf{61.67} & \textbf{29.91} & \textbf{26.34} & \textbf{23.70} & 80.15 & 78.51 & \textbf{58.87} & \textbf{59.72} & \textbf{60.30} & \textbf{48.08} \vspace{-0.3em}\\ 
& {\small±0.70} & {\small±0.00} & {\small±0.03} & {\small±0.34} & {\small±0.52} & {\small±0.01} & {\small±0.65} & {\small±0.65} & {\small±2.70} & {\small±0.30} & {\small±0.04} & {\small±0.90} \\
\hline
SPECTER & 61.85 & 65.62 & 64.20   & 27.57 & 21.75 & 17.43 & 78.74 & 78.18 & 53.75 & 56.05 & 55.18 & 45.13\\
{+FaBle} & \textbf{64.36} & \textbf{65.62} & \textbf{64.40} & \textbf{28.10} & \textbf{22.61} & \textbf{19.35} & \textbf{80.43} & \textbf{78.35} & \textbf{55.66} & \textbf{57.63} & \textbf{55.53} & \textbf{46.47}\vspace{-0.3em}\\
& {\small±0.63} & {\small±0.00} & {\small±0.02} & {\small±0.10} & {\small±0.00} & {\small±0.08} & {\small±0.00} & {\small±0.00} & {\small±0.02} & {\small±0.19} & {\small±0.00} & {\small±0.02} \\
\hline
\end{tabular}}\caption{Evaluation results on FEIR test set. FaBle denotes fine-tuning the SPECTER with our augmented dataset.}\label{table 4}
\end{table*}

\begin{table}[]
\centering
\scalebox{0.6}{%
\begin{tabular}{c|p{11.3cm}}
\hline
  &  \textbf{Question} \\ \hline
 Orig & \textbf{Why does the professor} ask the man to \textbf{come to her office}? \\
\hline
\rowcolor{cyan!5} \multirow{2}{*}{Sim} & What would be an appropriate \textbf{reason why the professor} might \textbf{invite} the student \textbf{to her office}? \\ 
\hline
\rowcolor{red!5} Dis &  What are some benefits of studying abroad? \\
\hline
    & \textbf{Options} \\ 
\hline
\multirow{4}{*}{Orig} & 1.The effect of the decrease in \textbf{temperatures} on wetlands\\
& 2.The use of computer models to analyze \textbf{temperature} patterns\\
& 3.The theory that land development \textbf{affected} the \textbf{climate} of South Florida \\
& 4.The importance of the citrus industry to the South Florida economy \\
\hline
\rowcolor{cyan!5}  & 1.The impact of urbanization on local \textbf{ecosystems} \\
\rowcolor{cyan!5} & 2.The role of water management practices in shaping \textbf{regional climates} \\
\rowcolor{cyan!5} & 3.The \textbf{influence} of agricultural activities on atmospheric conditions \\
\rowcolor{cyan!5} \multirow{-4}{*}{Sim} & 4.The effects of deforestation on biodiversity and \textbf{climate} \\
\hline
\rowcolor{red!5} & 1.The impact of social media on teenagers' self-esteem  \\
\rowcolor{red!5} & 2.The benefits of meditation for mental health \\
\rowcolor{red!5} & 3.The history of the civil rights movement in the United States \\
\rowcolor{red!5} \multirow{-4}{*}{Dis} & 4.The role of parental involvement in student academic achievement  \\
\hline
\end{tabular}
}\caption{Generated \textit{Sim} and \textit{Dis} facets of \textit{Question} and \textit{Options}. Relevant terms are highlighted in \textbf{bold}. 
}\label{table 6} 
\end{table}

In the \textit{result} facet, the impact of FaBle is evident, although not as large as the \textit{method}. This outcome aligns with prior models' performance, falling between the other two facets. Some \textit{result}s can be easily identified as similar by common phrase overlaps, while others demand a detailed interpretation of the query \cite{mysore2021csfcube}, where our sophisticated processing can be effective. The \textit{result}s are contextually dependent on other facets, as they typically discuss \textit{method}-driven observations or \textit{background}-posed problem-solving. Consequently, their similarities are shaped by overall abstract relevance \cite{mysore-etal-2022-multi}. The superiority of multi-match-based OTASPIRE over single-match-based TSASPIRE in the \textit{result} facet supports this. Thus, enriching FaBle with auxiliary methods addressing global-level similarity can be beneficial. 


\section{Analysis and Discussion}
\paragraph{{Impact of Hard Negatives}} 
We investigate the impact of hard-negative generation (§~\ref{sec3.5}). Before analyzing, we examine how our hard-negative sampling altered the score distribution of existing negatives. Figure~\ref{fig: figure 5} exhibits that regeneration shifted the average to around 0.75 points, aligning with our goal of acquiring more challenging samples. We only select samples below 0.5 as hard negatives to differentiate from positives. Table~\ref{table 1} (FaBle$_{\texttt{Spec}}$+HN) indicate that hard negatives for a specific facet, regenerated to have a higher similarity score, remarkably assist \textit{method}-faceted retrieval but not in the others. Creating high-similarity negative samples to a specific facet may hinder the relevance recognition on the general facets like \textit{background} and \textit{result}. Yet, for facets demanding a fine-grained approach, auxiliary optimizing with hard negatives can boost contrastive learning \cite{qu-etal-2021-rocketqa, santhanam-etal-2022-colbertv2, ostendorff-etal-2022-neighborhood, formal2022distillation}.




\paragraph{Comparison with Random Sampling} 
We compared the efficacy of directly generating negative facets to random sampling (Table~\ref{table 3}). In particular, we replaced dissimilar facet fragments created for each document with randomly selected original facets from other documents. Original facets are not defined in the document; thus, we utilize the summarized facets from Stage 1. Table~\ref{table 3} indicate that FaBle, which integrates generating \textit{dissimilar} component as specific-facet-tailored negatives, achieves markedly better performance than FaBle-RN, which employs random sampling. Hence, our subtle negative sampling may be a key for faceted QBE, aligning with contemporary research that emphasizes the advantages of strategic negative sampling over random approaches \cite{qu-etal-2021-rocketqa, zhan2021optimizing, zhou-etal-2023-towards-robust}.

\paragraph{Effects of the Data Size} 
We examine how the amount of augmentation affects model performance. For 50\%, we randomly select half the original document (0.5K out of 1K), creating 20K triplet pairs with FaBle. Figure~\ref{fig: figure 7} reveals that increasing the data size consistently enhances NDCG\%20 and MAP scores. For both metrics, the \textit{Background} facet shows reasonably high scores even at the base level, implying that the model itself could represent this comprehensible facet well; hence, fine-tuning on larger data moderately impacts the model performance. Meanwhile, the \textit{Method} facet, indicated to be underrepresented in the baseline model by exhibiting lagged performance behind the other two facets, shows a clear improving tendency as the amount of FaBle-augmented data increases. Thus, tailoring data size to the specific needs of individual facets is essential for training optimization. 


\paragraph{Results on FEIR}
Table~\ref{table 4} presents the experimental results of fine-tuning SPECTER on FaBle-augmented data with educational items. FaBle brings in performance improvements in all facets, with the substantial 3.6\% NDCG and 3.4\% MAP increases in the \textit{question} facet, mirroring trends of the CSFCube results (Table~\ref{table 2}). Generally, when holistically finding similar items using a coarse-grained approach, \textit{question}, which comprises a single sentence, is more likely to be overlooked than options' four sentences and a story of a paragraph constituting multiple sentences. In contrast, FaBle constructs facet-specific positive and negative documents with modularized combinations, allowing even less prominent facets to be targeted.
In the qualitative analysis for generation (Table~\ref{table 6}), similar components are content-relevant to the original, while dissimilar ones shift to irrelevant topics, implying FaBle's ability to fit intentions.
\section{Conclusion}
We introduce FaBle, a multi-facet blending augmentation that aids in direct model training for faceted QBE. By modularizing facets by decomposing and recomposing, FaBle effectively synthesizes pseudo-documents that match user-intended facets, eliminating the need for pre-set annotations. FaBle improves the retrieval performances, particularly in the salient facet, surpassing models trained on much larger datasets. In addition, we release a FEIR test set for the language education domain, demonstrating FaBle's generalizability. 

\section*{Limitations}
Currently, we assume data scarcity by applying FaBle on a small amount of data to evaluate the assistance in real-world settings where open corpora are limited. However, as we observed the performance-improving trends with the increased amount of datasets, augmenting with more original data could lead to further enhancements. Secondly, in prior faceted QBE works, statistical tests are not provided,
which may be attributed to the small test set size (e.g., 16-17 queries per facet in CSFCube) confining statistical power. Nevertheless, to investigate the robustness of FaBle, we examined the proportion of queries where performance remained equal or improved. Among the aggregated queries, 70.83\%, 70.83\%, and 75\% showed increased NDCG$_{\%20}$, NDCG$_{\%10}$, and MAP scores over SPECTER, respectively, demonstrating that FaBle is effective for the majority of queries.

\section*{Ethical Statement}
We follow the guidelines outlined in the ACL Code of Ethics. Our work did not utilize private datasets, and it does not include any confidential personal information. For the human annotation, we provide fair compensation for two human experts, paying \$135 as a fixed price.

\section*{Acknowledgements}

This research was partly supported by the MSIT(Ministry of Science and ICT), Korea, under the ITRC(Information Technology Research Center) support program(IITP-2024-2020-0-01789) supervised by the IITP(Institute for Information \& Communications Technology Planning \& Evaluation), and by Institute of Information \& communications Technology Planning \& Evaluation (IITP) grant funded by the Korea government(MSIT) (No.2022-0-00223, Development of digital therapeutics to improve communication ability of autism spectrum disorder patients).

\bibliography{anthology, main}

\appendix

\section{Relevance Annotation for FEIR}
\label{appendix3}
For FEIR relevance annotation, experts rated the relevance degree of each query and candidate item within the facet on an integer scale from 0 to 3, similar to \citet{mysore2021csfcube}, according to the following guidelines:
\begin{itemize}
  \item 3 (Near Identical): A strong and clear correlation exists between the facet of a query item and a candidate item. Significant overlap in content, background, context, or theme indicates a high association level.
  \item 2 (Similar): An apparent degree of connection is observed between the facet of a query item and a candidate item. Shared elements or themes suggest a moderate level of association.
  \item 1 (Related): A superficial connection exists but is minimal. There may be slight thematic or contextual similarities, but the items are mainly independent.
  \item 0 (None or Irrelevant): Items that do not meet the criteria for the above categories should be labeled as 0.
\end{itemize}

\section{Experimental Settings}
\label{appendix1}
For all the experiments, we report the average results conducted by three runs with different seeds, \texttt{\small\{22,222,2222\}}. The batch size is set as 30, following the settings of the previous model \citep{mysore-etal-2022-multi}. 
We use a 1e-5 learning rate and two epochs for academic paper retrieval and a 1e-6 learning rate and two epochs for education items. Margin $m$ for the triplet loss is set as 1. Fine-tuning and model inference are performed using an A100-SXM4-40GB GPU and take approximately 2 hours. For LLaMA, we used the LLaMA2-13B chat model\footnote{https://ai.meta.com/llama/}. To fine-tune SPECTER, we split the generated dataset into training and validation sets with a 9:1 ratio. 


\section{Baseline Models}
~\label{appendix2}
We examine the comparability of our method's performance with competitive models that exhibited strong results in facet-conditional retrieval. In particular, we report the results of a robust faceted QBE model, ASPIRE \citep{mysore-etal-2022-multi}, and its various comparisons, outlined in \citet{mysore2021csfcube} and \citet{mysore-etal-2022-multi}. The reported TSASPIRE$_{\texttt{Spec}}$ is a SPECTER-based single-match method with textual supervision, OTASPIRE$_{\texttt{Spec}}$ is a multi-match method utilizing optimal transport, and TS+OTASPIRE$_{\texttt{Spec}}$ combines both approaches, as a multi-task and multi-aspect method. They are trained with 1.3M training sets. Their comparison models, \texttt{SentBERT-PP}, \texttt{SentBERT-NLI}, and \texttt{CoSentBert}, are MPNET-1B\footnote{MPNET-1B is pre-trained over 1B text pairs} based sentence embedding models. SPECTER-ID results are also reported, fine-tuned with 660K in-domain papers that fit the CSFCube test set.

\section{Evaluation Metrics}
For evaluation, we employ normalized discounted cumulative gain at rank K (NDCG@K) and mean average precision (MAP), well-known retrieval metrics. In particular, we set $K=p*|C|$ where $p \in (0,1)$ and report NDCG$_{\%20}$, which denotes computing at 20\% of the query pool size, following existing research \cite{wang2013theoretical, mysore2021csfcube, mysore-etal-2022-multi}. The NDCG metric reflects the graded relevance scores of items to assess the ranking quality, offering a more nuanced perspective than binary metrics such as precision or recall, particularly when the dataset is annotated with multiple relevance scores. Given that our test sets, CSFCube and FEIR, have multiple numeric relevance annotations, NDCG would be the most suitable metric. Specifically, for the FEIR test set, which has fewer queries and candidates than CSFCube, we also report the NDCG$_{\%10}$ results, which compute at 10\% of the query pool size.


\end{document}